\documentclass[rnote]{aa}  

\usepackage{natbib}
\bibpunct{(}{)}{;}{a}{}{,} 

\usepackage{graphicx}
\usepackage[varg]{txfonts}
\begin{document}

   \title{Two barium stars in the Galactic bulge\thanks{Based on observations at the Very 
   Large Telescope of the European Southern Observatory, Cerro Paranal/Chile, under
   Programme 083.D-0046(A)}}

   \author{T. Lebzelter
          \inst{1}
          \and
          S. Uttenthaler
          \inst{1}
          \and
          O. Straniero
          \inst{2}
          \and
          B. Aringer
          \inst{1}
          }

   \institute{ University of Vienna, Department of Astrophysics,
              T\"urkenschanzstrasse 17, A-1180 Vienna\\
              \email{[thomas.lebzelter;stefan.uttenthaler]@univie.ac.at}
              \and
              INAF, Osservatorio Astronomico di Collurania, 64100 Teramo, Italy\\
              \email{straniero@oa-teramo.inaf.it}
             }

   \date{Received ; accepted }

 
  \abstract
   {Barium stars conserve important information on the s-process and the third dredge-up in intermediate
   mass stars. Their discovery in various environments is therefore of great help in testing nucleosynthesis
   and mixing models.}
   {Our aim is to analyse two stars with a very strong barium line detected in a large survey of red giants
   in the Galactic bulge.}
   {Abundance analysis was done comparing synthetic model spectra based on the COMARCS code with our
   medium resolution spectra. Abundances of Ba, La, Y, and Fe were determined.
   Beside the two main targets, the analysis was also applied to two comparison
   stars.}
   {We confirm that both stars are barium stars. They are the first ones of this kind identified in the
   Galactic bulge. Their barium excesses are among the largest values found up to now. The elemental
   abundances are compared with current nucleosynthesis and mixing models. Furthermore, we estimate a frequency of
   barium stars in the Galactic bulge of about 1\,\%, which is identical to the value for disc stars.}
   {}


   \maketitle
%

\section{Introduction}
Low- and intermediate-mass stars manufacture heavy elements via the s-process during their final evolutionary
stage, the so-called thermal pulse asymptotic giant branch (TP-AGB). Strong enrichment of these elements
in an object therefore normally indicates that the star is already in this phase. However, there is a group
of stars showing enhancement in s-process elements while being far away from the TP-AGB phase. Because of the
prominent lines of the s-process element barium in their spectra, a sub-group of these objects is called barium-stars 
\citep{1951ApJ...114..473B}. These stars also show unusually strong bands of CH and CN. The most likely
interpretation of these objects is a mass-transfer scenario in a binary system. The more massive companion has
already evolved through the TP-AGB phase and deposited material enriched by nucleosynthesis on the less
evolved companion, which now appears as a barium star \citep{1988A&A...205..155B}.

Barium stars are interesting objects since they preserve the abundance signature of more massive AGB stars,
which is of particular importance for studying metal-poor objects. Their abundance pattern is widely used 
in calibrating models for s-process nucleosynthesis on the AGB \citep{1999ARA&A..37..239B,2001ApJ...557..802B}.
They also give us insight into the mass transfer process in binaries and envelope mixing. 

A few hundred barium stars are known in the Milky Way, most of them belonging to the disc population, but some 
show characteristics of halo kinematics \citep{1997A&A...319..881G}. The detection of these objects in various
environments allows them to be used as probes for AGB nucleosynthesis at various metallicities \citep{2011A&A...533A..51P}. The excess of s-process elements in barium stars is typically found
in the range +0.6$<$[X/Fe]$<$+1.8 \citep{2006A&A...454..895A} with an increase in the overabundances
of the heavy s-elements with decreasing metallicity \citep{2006A&A...454..895A,2011A&A...533A..51P}.
This can be understood in terms of an increased ratio of seed nuclei to free neutrons 
at higher metallicity and is in agreement
with predictions \citep{1999ARA&A..37..239B}.

In this paper we report the detection of two additional barium stars. Remarkably, they are located in the
Galactic bulge, where no barium stars have been reported yet (see below for some candidates). 
   
\section{Methods}
\subsection{Observations}
The detection of two barium stars appeared as a by-product of our spectroscopic study on a sample of
400 red giants in the Palomar-Gronigen field \# 3 (PG3) 
of the Galactic bulge. Results on the lithium content have been 
published in \citet[][hereafter Paper I]{2012A&A...538A..36L}. 
The same sample has also been used for a detailed study on
kinematics and metallicity distribution of red giants in the bulge, see \citet[][hereafter Paper II]{2012A&A...546A..57U}.
Observation of the sample and data reduction have been described in detail in these two papers. We therefore
repeat only the basic numbers here.

\begin{figure}
\resizebox{\hsize}{!}{\includegraphics{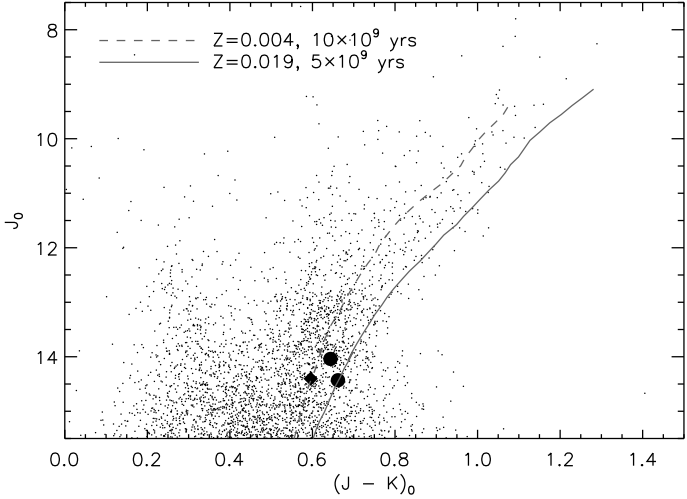}}
\caption{Dereddened 2MASS colour-magnitude diagram of the barium stars (filled diamond symbols) and the comparison 
stars (filled circles). The barium star ID 410 basically coincides with its comparison star ID 409, which is why only 
the comparison star (the brighter one) is visible here. All other stars shown here are within a 12.5 arcmin diameter 
field around (l,b)=(0\,$^{\circ}$,$-$10\,$^{\circ}$). 
The solid and dashed grey lines are isochrones from \citet{2000A&AS..141..371G} 
with the metallicities and ages as indicated in the legend. They were used to select spectroscopic targets in the 
bulge (see Paper I) and are truncated at the RGB tip. The RGB bump of these isochrones is at $J_0\approx14\fm0$.}
\label{cmd}
\end{figure}

The observations were done using the FLAMES-GIRAFFE spectrograph at ESO's VLT. A colour-magnitude diagram
(CMD) of the complete sample has been presented in Fig. 1 of Paper I. The spectra
cover the range between 644 and 682\,nm at a resolution of 17000. The two barium stars reported here were
discovered by visual inspection of the spectra, because they showed an outstanding absorption line near 6500\,{\AA}. 
Both stars are found at the bottom of the red giant branch. Basic stellar parameters taken from
Paper II are summarized in Table\,\ref{basics}. The absolute bolometric magnitude was derived from the 
$K$ magnitude with a bolometric correction term taken from Paper I for a bulge distance modulus of 14.5.
We get absolute bolometric luminosities $M_{\rm bol}$ between 1$\fm$2 and 1$\fm$6 which is about 5 magnitudes
less than the third dredge-up limit \citep[cf.][]{2000A&AS..141..371G,2007A&A...463..251U}. Our
stars are found at a brightness close to the less luminous red clump in the bulge \citep[cf.][]{2012A&A...546A..57U}.
We note, however, that owing to the extension of the bulge and the evolutionary path passing twice through a similar
point in the Hertzsprung-Russell-diagram, it is not possible to unambiguously attribute any of our programme 
stars to that clump. Figure \ref{cmd} shows the location of the two barium stars in a colour-magnitude diagram of the surrounding bulge field.

Since we have only a single spectrum for each star from our programme, we cannot investigate radial velocity changes 
as an indication for a possible binary nature. Our field was also observed within the ARGOS survey 
\citep{2013MNRAS.428.3660F}, but unfortunately the barium star candidates were not observed in that survey (M. Ness, 
private comm.). For comparison, we analysed two more stars in our sample with similar stellar parameters 
but with no indication of an increased Ba abundance. Their location in a colour-magnitude diagram of the bulge
field is visible in Fig.~\ref{cmd}.
One of these comparison stars (star 497) was observed by ARGOS, and 
parameters very similar to our results were found, including the radial velocity ($-30.8$\,km\,s$^{-1}$).

\begin{table*}
\caption{Basic parameters of the analysed targets.} 
\label{basics} 
\centering 
\begin{tabular}{lrrrrrrrrrrl} 
\hline\hline 
ID & RA (2000) & Dec (2000) & $K$ & $J-K$ & T$_{\rm eff}$ & RV$_{\rm helio}$ & [Fe/H] & [$\alpha$/Fe] & B.C. & $M_{\rm bol}$ & comment\\
 & & & [mag] & [mag] & [K] & [km\,s$^{-1}$] &  & & & &\\
\hline 
410 &  18:27:00.8 & -33:44:18 & 13.448 & 0.725 & 4659 & +24.1 & $-$0.28 & +0.23 & 2.2 & 1.2 & barium star\\
490 &  18:26:50.9 & -33:47:39 & 13.856 & 0.675 & 4810 & $-$2.5 & $-$0.56 & +0.23 & 2.2 & 1.5 & barium star\\
409 &  18:26:18.9 & -33:42:21 & 13.446 & 0.725 & 4665 & +0.5 & $-$0.04 & +0.14 & 2.2 & 1.2 & comparison star\\
497 &  18:27:19.4 & -33:53:46 & 13.816 & 0.739 & 4612 & $-$30.9 & $-$0.52 & +0.20 & 2.3 & 1.6 & comparison star\\
\hline 
\end{tabular}
\tablefoot{All data are taken from Paper II. $M_{\rm bol}$ calculated using (m-M)=14.5 and a bolometric correction
(B.C.) as in Paper I.}
\end{table*}

\subsection{Abundance determination}
As in the previous papers of this series, we used a comparison of observed and synthetic spectra to determine the
abundances of selected elements in our two target stars. To compute the synthetic spectra, model atmospheres calculated with the COMARCS code \citep{2009A&A...503..913A} were used. The effective temperatures were set to
4660\,K, 4810\,K, 4660\,K, and 4610\,K, respectively (see Table\,\ref{basics}). For all four stars a log\,g value of
2.5 was used. The microturbulence parameter was set to 1.5\,km\,s$^{-1}$, which was found to give the best
overall fit of the observed spectra. 
The synthetic spectra were computed with a resolution of R$=$200000 and then rebinned
to 17000. A macroturbulence broadening of 3\,km\,s$^{-1}$ was applied.

The metallicity determined in Paper II (Table\,\ref{basics})
was taken as a starting value which was further refined in the present
analysis. The goal was to determine abundances for Y, Ba, and La. Because of the limited spectral range covered by
our observations and the severe line blending, the determination could be based on only one line of these elements. 
For Ba this was
the line at 6498.7\,{\AA}, for La the line at 6528.9\,{\AA}, and for Y the blend with an iron line at 6615\,{\AA}.
All three lines stem from transitions of single ionized atoms. Both the Ba and the La line show a hyperfine structure. This was included in the analysis using the data given in \citet{1978SoPh...56..237R} for Ba and
in \citet{2001ApJ...556..452L} for La, respectively. Calculations were done in LTE following the calculations
of \citet{2006ApJ...641..494S} showing a very small NLTE effect on the abundance derived from the Ba line chosen.
The solar abundance pattern was taken from \citet{2008A&A...488.1031C}.

For the two barium stars the abundances of all other
s-process elements were increased by 0.5 dex in the spectrum calculations, but because of the lack of unblended lines
and the limits set by the S/N in these comparably faint stars of our sample no attempt was made to fit further
element abundances. 
Contrary to the work done in Paper II which relied primarily on an automatic fitting approach, 
here the final decision on the best fit was done by visual comparison of model and observation. 
The clearly visible CN features in the two barium stars required a slight increase in the N 
abundance\footnote{In principle, the strength of the CN bands can also be altered by an increase in the C
abundance or by a moderate increase in both abundances. While the best fit was achieved by changing the N abundance,
the other two options cannot be ruled out completely.} 
which is in agreement with 
expections for this kind of objects. In the analysis of Papers I and II the abundance of N was not altered within
the sample. The increased N abundance in these two objects was then obviously 
compensated by a higher Fe abundance in the automatic
analysis. Therefore, we had to make a refinement of the Fe abundance for these two objects in our present study.
However, it seems that this problem affected 
the metallicity determination in only these two stars, since our analysis of the comparison stars gave the same
Fe abundance as the automatic analysis in Paper II.
 
To estimate the uncertainties of the derived abundances we calculated additional synthetic spectra varying
log\,g and T$_{\rm eff}$ by $\pm$0.5 and by $\pm$100\,K, respectively.
Both ranges are the maximum uncertainties in these parameters expected from the analysis in Paper II, and the 
resulting errors are certainly on the side of caution. In addition, the uncertainty in the exact location 
of the pseudo-continuum level has been taken into account. The three errors were then combined quadratically.
The resulting uncertainties of [X/Fe] given in Table \ref{abundances} also include the uncertainty in metallicity.
Note that the abundances are based on the fit of a single line only. The comparably large error in the
Y abundance is a result of the studied line being very sensitive to changes in log\,g.
Uncertainties in the N abundance, as well as uncertainties in the solar reference values, 
were not considered because of
their small effect on the result. We also did a test using a higher C/O ratio of 0.9 in the model, 
which may be more appropriate for a highly evolved AGB star transfering mass on the companion. This change is
hardly visible in the model spectrum in the selected wavelength region, and thus has no effect on the derived
s-process element abundances.

\section{Results}

\begin{figure}
\resizebox{\hsize}{!}{\includegraphics{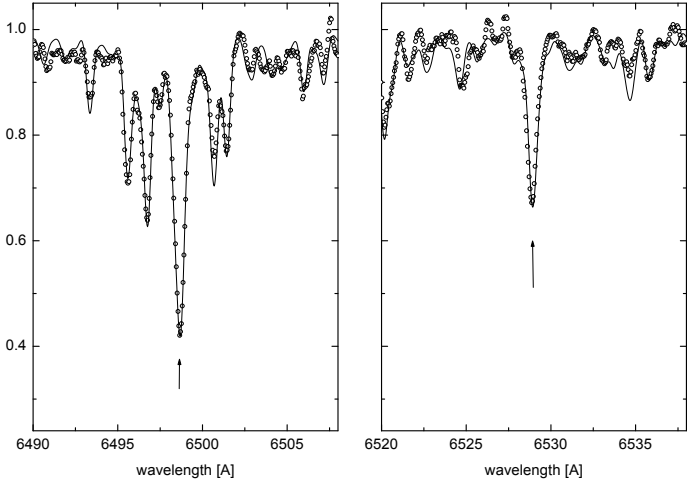}}
\caption{\ion{Ba}{ii} line (left) and \ion{La}{ii} line (right, marked by an arrow) used to derive
the corresponding element abundances. The observed spectrum of star 490 is shown by open symbols;
the solid line gives the best fit model.}
\label{barLan}
\end{figure}

Figure \ref{barLan} shows an example of the model fit of the observed spectra. The overall agreement is
satisfying. The results are summarized in Table \ref{abundances}. The elments 
Ba and La are strongly enhanced in both 
barium stars. A less expressed enhancement is also seen in Y with a larger error bar. As mentioned above
the iron abundance is slightly different from the values derived in Paper II. We also note that the 
fit of the \ion{Sc}{ii} line at 6606.5\,{\AA} can be
improved when reducing the abundance of this element by about 0.4 dex in these objects. 
From Paper I we know that the stars do not
show any significant abundance of Li.

The analysis also confirmed that the two comparison stars were not barium stars. None of the tested
elements shows any considerable enhancement above the scaled solar value in these cases.

\begin{table*}
\caption{Derived abundances for selected elements.} 
\label{abundances} 
\centering 
\begin{tabular}{lrrrrrrrr} 
\hline\hline 
ID & [Fe/H] & $\Delta$[Fe/H] & [Ba/Fe] & $\Delta$[Ba/Fe] & [La/Fe] & $\Delta$[La/Fe] & [Y/Fe] & $\Delta$[Y/Fe]\\
\hline 
410 & $-$0.5 & 0.2 & +2.3 & 0.4 & +1.8 & 0.5 & 0.9 & 0.5\\
490 & $-$0.7 & 0.1 & +1.8 & 0.2 & +1.5 & 0.3 & 0.8 & 0.5\\
409 & $-$0.1 & 0.2 & $-$0.1 & 0.4 & 0.0 & 0.4 & 0.0 & 0.6\\
497 & $-$0.5 & 0.2 & $-$0.2 & 0.4 & 0.0 & 0.4 & 0.0 & 0.6\\
\hline
\end{tabular}
\end{table*}

In both stars 410 and 490, barium is more enhanced than lanthanum, although the difference is within the 
error bars. A key indicator used for the comparison of TP-AGB models with observations is the ratio
between the light and the heavy s-process elements. Classicaly, this ratio is formed using the Y and Zr abundance
on the one hand and the La, Nd, and Sm abundance on the other \citep[e.g.][]{2009PASA...26..176H}.
We can only give a simplified version here forming the ratio between [Y/Fe] and [La/Fe]. This gives 
for [hs/ls] values of 0.9 and 0.7 for star 410 and star 490, respectively. The considerable uncertainty
in both abundances limits the reliability of this result. However, errors in temperature and log\,g alter
the abundances of La and Y in the same direction, thus giving a combined uncertainty of only about 0.3 dex.

\section{Discussion}
To our knowledge, this is one of the first detections of Ba stars in the Galactic bulge. \citet{2012ApJ...749..175J}
detected one giant in their sample of 69 bulge stars with a remarkable overabundance of [La/Fe]$=$+1.43$\pm$0.32,
and [Fe/H]$=-$0.56. That star is also enriched in Zr and Nd. However, \citet{2012ApJ...749..175J} did not discuss
this object further. Using the star's $K$ magnitude and the same approach as for our targets, we get an $M_{\rm bol}$=$-$0.9. 
This star (2MASS 18174742-3348098) is, thus, much brighter than our targets but still less luminous than 
the 3DUP limit. We therefore consider this object a barium star as well.
\citet{2003SPIE.4834...66C} found enhancements of several s-process elements in a few bulge
stars studied using a microlensing technique. Among them is an interesting dwarf star, 99-BLG-1, which shows
a strong enhancement in Ru, La, and Ce, but surprisingly not in Ba, which is solar in this object. We note, however,
that \citet{2003SPIE.4834...66C} did not apply a correction for hfs splitting of the analysed lines, 
therefore the actual error bar may be larger than their given value of 0.2 dex.  

With [Fe/H] of about $-$0.5, our two barium stars are located close to the metallicity where the maximum 
enhancement for the second s-process peak (Ba peak) is expected \citep[cf. Fig.12 in][]{1999ARA&A..37..239B}.
The measured overabundances for Ba and La agree well with the predictions for a 1.5\,M$_{\sun}$ TP-AGB star
of that metallicity with the standard choice of the $^{13}$C pocket efficiency \citep{2009PASA...26..176H}.
On the contrary, the measured abundances for Y are a bit lower than the model calculation. More recent calculations
by \citet{2011ApJS..197...17C} include a simultaneous solution of the AGB evolution of the physical structure and
the chemical evolution\footnote{http://fruity.oa-teramo.inaf.it/}. For the metallicity given in Table 
\ref{abundances} we find a good agreement of our [Ba/Fe] and [La/Fe] values for an AGB star around 2.0 to 
2.5\,M$_{\sun}$, which would be the companion of the barium star. 
The Y abundance predicted by these models agrees with our measurements within the error bars.
 
Compared to the measured excesses in literature studies on larger samples
\citep{2006A&A...454..895A, 2007A&A...468..679S, 2011A&A...533A..51P}, our stars are found in the top range 
for the Ba and La abundance excesses for all barium stars. In particular, star 410 may be the star with the 
highest Ba overabundance detected up to now. Compared to the Y abundance excesses found for barium stars of
that metallicity by \citet{2006A&A...454..895A}, the two objects studied here are found at the lower end of
the distribution.

The measured abundance of Y, Ba, and La in a barium star depends also on the dilution factor describing the
change of surface abundances due to mixing of the accreted material with the deeper layers of the 
original stellar atmosphere.
The large excess of Ba and La observed seems to be compatible with the hypothesis that no major mixing
event has occurred in the stars since the accretion took place. 
The derived log\,g value suggests that both our barium stars are indeed giants. We may speculate that the
mass transfer from the TP-AGB star happened after the major mixing event of the first dredge-up. According to
\citet{2000A&AS..141..371G}, the total life times of a star of 2 and 2.5\,M$_{\sun}$ are 1.1$\times$10$^{9}$ and 5.9$\times$10$^{8}$ years, respectively. If this hypothesis is correct and the accretion process occurred after the
first dredge-up of the companion, the mass of the barium star has 
to be chosen such that it reaches the RGB within
the total life time of the donor star. This would mean that our two barium stars, in particular star 410, 
have a mass larger than 
1.7\,M$_{\sun}$. Such a mass estimate would nicely agree with the existence of an intermediate population
in the bulge suggested, e.g. by \citet{2007A&A...463..251U} to explain the existence of the third dredge-up
indicator Tc in bulge AGB stars. However, this scenario is affected by the various uncertainties
in the data, so that a final conclusion cannot be drawn.

An alternative conclusion would be that the 
calculated AGB yields are underestimated. In that case, it would imply that the third dredge-up is 
deeper than estimated in the calculations of, e.g. \citet{2011ApJS..197...17C}, and/or the mass loss rate
is lower than assumed. Concerning the mass loss rate, \citet{2009ApJ...696..797C} expect an increase in the
La yields of 24\,\% if the mass loss rate is reduced by a factor of 2. A more efficient third dredge-up, 
resulting from a larger mixing length parameter in the models, also leads to a significant increase of
the predicted La abundance. The fact that both of our barium stars come with a high Ba abundance may
suggest that the second scenario, an underestimate of AGB yields by the models, is more probable since
the probability that both stars are found immediately after the mass transfer from the companion is
expected to be rather low. However, it is remarkable that there are not more stars known with a barium
abundance similar to star 410. 

Because of the surprisingly low abundance of Y, the ratio [hs/ls] is clearly higher in our two barium stars than
the value expected from the models \citep[cf. Fig. 7 in][]{2001ApJ...557..802B}. However, it is not completely
out of the range of observed values, and considering the error bar it may well fit within the range of the models.
An extended study based on more lines and including an abundance analysis of further elements from the Y/Zr-peak
would be needed to give a proper [hs/ls] ratio. Its obvious advantage for a comparison with stellar nucleosynthesis
models would be its independency of the 
uncertainties of dredge-up efficiency, mass loss rate, and dilution factor. 

The two Ba stars detected here and the one from \citet{2012ApJ...749..175J} are probably members of the metal-poor population identified previously in the bulge \citep[Paper II,][]{2011A&A...534A..80H}. 
In Paper II we found that the transition between the metal-poor and metal-rich population is at ${\rm 
[Fe/H]}\sim-0.20$, meaning that 243 giants from that sample belong to the metal-poor population. In Johnson's 
paper, 31 bulge RGB stars are more metal-poor than this limit, one of which is the Ba star discussed above. This 
means that in the metal-poor population three out of 274 giants are 
Ba stars, giving a Ba-star fraction of $\sim1$\%. This is identical with the findings from the disc field
\citep{2010A&A...523A..10I, 1972AJ.....77..384M}.
Ba stars in the metal-rich bulge population yet remain to be identified. It is thus possible that the fraction of 
binaries that allow for the formation of Ba stars in that population is lower than in the metal-poor population and 
the disc. This would be in agreement with expectations since metal poor stars are known to show a more efficient
dredge-up during the AGB phase.
However, at the moment there are only very few studies of the s-element content of 
bulge stars available.

\begin{acknowledgements}
      The work of TL has been supported by the Austrian Science Fund
      under project number P23737-N16. SU acknowledges support from the Austrian Science Fund (FWF) 
      under project P~22911-N16. The authors thank Martin Stift for helpful discussions on the hyperfine 
      structure of lines, and Melissa Ness for 
      providing the ARGOS results for one of the comparison stars.
\end{acknowledgements}

\bibliographystyle{aa} 
\bibliography{bariumrefs} 

\begin{thebibliography}{26}
\expandafter\ifx\csname natexlab\endcsname\relax\def\natexlab#1{#1}\fi

\bibitem[{{Allen} \& {Barbuy}(2006)}]{2006A&A...454..895A}
{Allen}, D.~M. \& {Barbuy}, B. 2006, \aap, 454, 895

\bibitem[{{Aringer} {et~al.}(2009){Aringer}, {Girardi}, {Nowotny}, {Marigo}, \&
  {Lederer}}]{2009A&A...503..913A}
{Aringer}, B., {Girardi}, L., {Nowotny}, W., {Marigo}, P., \& {Lederer}, M.~T.
  2009, \aap, 503, 913

\bibitem[{{Bidelman} \& {Keenan}(1951)}]{1951ApJ...114..473B}
{Bidelman}, W.~P. \& {Keenan}, P.~C. 1951, \apj, 114, 473

\bibitem[{{Boffin} \& {Jorissen}(1988)}]{1988A&A...205..155B}
{Boffin}, H.~M.~J. \& {Jorissen}, A. 1988, \aap, 205, 155

\bibitem[{{Busso} {et~al.}(2001){Busso}, {Gallino}, {Lambert}, {Travaglio}, \&
  {Smith}}]{2001ApJ...557..802B}
{Busso}, M., {Gallino}, R., {Lambert}, D.~L., {Travaglio}, C., \& {Smith},
  V.~V. 2001, \apj, 557, 802

\bibitem[{{Busso} {et~al.}(1999){Busso}, {Gallino}, \&
  {Wasserburg}}]{1999ARA&A..37..239B}
{Busso}, M., {Gallino}, R., \& {Wasserburg}, G.~J. 1999, \araa, 37, 239

\bibitem[{{Caffau} {et~al.}(2008){Caffau}, {Ludwig}, {Steffen}, {Ayres},
  {Bonifacio}, {Cayrel}, {Freytag}, \& {Plez}}]{2008A&A...488.1031C}
{Caffau}, E., {Ludwig}, H.-G., {Steffen}, M., {et~al.} 2008, \aap, 488, 1031

\bibitem[{{Cavallo} {et~al.}(2003){Cavallo}, {Cook}, {Minniti}, \&
  {Vandehei}}]{2003SPIE.4834...66C}
{Cavallo}, R.~M., {Cook}, K.~H., {Minniti}, D., \& {Vandehei}, T. 2003, in
  Society of Photo-Optical Instrumentation Engineers (SPIE) Conference Series,
  Vol. 4834, Society of Photo-Optical Instrumentation Engineers (SPIE)
  Conference Series, ed. P.~{Guhathakurta}, 66--73

\bibitem[{{Cristallo} {et~al.}(2011){Cristallo}, {Piersanti}, {Straniero},
  {Gallino}, {Dom{\'{\i}}nguez}, {Abia}, {Di Rico}, {Quintini}, \&
  {Bisterzo}}]{2011ApJS..197...17C}
{Cristallo}, S., {Piersanti}, L., {Straniero}, O., {et~al.} 2011, \apjs, 197,
  17

\bibitem[{{Cristallo} {et~al.}(2009){Cristallo}, {Straniero}, {Gallino},
  {Piersanti}, {Dom{\'{\i}}nguez}, \& {Lederer}}]{2009ApJ...696..797C}
{Cristallo}, S., {Straniero}, O., {Gallino}, R., {et~al.} 2009, \apj, 696, 797

\bibitem[{{Freeman} {et~al.}(2013){Freeman}, {Ness}, {Wylie-de-Boer},
  {Athanassoula}, {Bland-Hawthorn}, {Asplund}, {Lewis}, {Yong}, {Lane}, {Kiss},
  \& {Ibata}}]{2013MNRAS.428.3660F}
{Freeman}, K., {Ness}, M., {Wylie-de-Boer}, E., {et~al.} 2013, \mnras, 428,
  3660

\bibitem[{{Girardi} {et~al.}(2000){Girardi}, {Bressan}, {Bertelli}, \&
  {Chiosi}}]{2000A&AS..141..371G}
{Girardi}, L., {Bressan}, A., {Bertelli}, G., \& {Chiosi}, C. 2000, \aaps, 141,
  371

\bibitem[{{Gomez} {et~al.}(1997){Gomez}, {Luri}, {Grenier}, {Prevot},
  {Mennessier}, {Figueras}, \& {Torra}}]{1997A&A...319..881G}
{Gomez}, A.~E., {Luri}, X., {Grenier}, S., {et~al.} 1997, \aap, 319, 881

\bibitem[{{Hill} {et~al.}(2011){Hill}, {Lecureur}, {G{\'o}mez}, {Zoccali},
  {Schultheis}, {Babusiaux}, {Royer}, {Barbuy}, {Arenou}, {Minniti}, \&
  {Ortolani}}]{2011A&A...534A..80H}
{Hill}, V., {Lecureur}, A., {G{\'o}mez}, A., {et~al.} 2011, \aap, 534, A80

\bibitem[{{Husti} {et~al.}(2009){Husti}, {Gallino}, {Bisterzo}, {Straniero}, \&
  {Cristallo}}]{2009PASA...26..176H}
{Husti}, L., {Gallino}, R., {Bisterzo}, S., {Straniero}, O., \& {Cristallo}, S.
  2009, \pasa, 26, 176

\bibitem[{{Izzard} {et~al.}(2010){Izzard}, {Dermine}, \&
  {Church}}]{2010A&A...523A..10I}
{Izzard}, R.~G., {Dermine}, T., \& {Church}, R.~P. 2010, \aap, 523, A10

\bibitem[{{Johnson} {et~al.}(2012){Johnson}, {Rich}, {Kobayashi}, \&
  {Fulbright}}]{2012ApJ...749..175J}
{Johnson}, C.~I., {Rich}, R.~M., {Kobayashi}, C., \& {Fulbright}, J.~P. 2012,
  \apj, 749, 175

\bibitem[{{Lawler} {et~al.}(2001){Lawler}, {Bonvallet}, \&
  {Sneden}}]{2001ApJ...556..452L}
{Lawler}, J.~E., {Bonvallet}, G., \& {Sneden}, C. 2001, \apj, 556, 452

\bibitem[{{Lebzelter} {et~al.}(2012){Lebzelter}, {Uttenthaler}, {Busso},
  {Schultheis}, \& {Aringer}}]{2012A&A...538A..36L}
{Lebzelter}, T., {Uttenthaler}, S., {Busso}, M., {Schultheis}, M., \&
  {Aringer}, B. 2012, \aap, 538, A36, {(Paper I)}

\bibitem[{{MacConnell} {et~al.}(1972){MacConnell}, {Frye}, \&
  {Upgren}}]{1972AJ.....77..384M}
{MacConnell}, D.~J., {Frye}, R.~L., \& {Upgren}, A.~R. 1972, \aj, 77, 384

\bibitem[{{Pereira} {et~al.}(2011){Pereira}, {Sales Silva}, {Chavero}, {Roig},
  \& {Jilinski}}]{2011A&A...533A..51P}
{Pereira}, C.~B., {Sales Silva}, J.~V., {Chavero}, C., {Roig}, F., \&
  {Jilinski}, E. 2011, \aap, 533, A51

\bibitem[{{Rutten}(1978)}]{1978SoPh...56..237R}
{Rutten}, R.~J. 1978, \solphys, 56, 237

\bibitem[{{Short} \& {Hauschildt}(2006)}]{2006ApJ...641..494S}
{Short}, C.~I. \& {Hauschildt}, P.~H. 2006, \apj, 641, 494

\bibitem[{{Smiljanic} {et~al.}(2007){Smiljanic}, {Porto de Mello}, \& {da
  Silva}}]{2007A&A...468..679S}
{Smiljanic}, R., {Porto de Mello}, G.~F., \& {da Silva}, L. 2007, \aap, 468,
  679

\bibitem[{{Uttenthaler} {et~al.}(2007){Uttenthaler}, {Hron}, {Lebzelter},
  {Busso}, {Schultheis}, \& {K{\"a}ufl}}]{2007A&A...463..251U}
{Uttenthaler}, S., {Hron}, J., {Lebzelter}, T., {et~al.} 2007, \aap, 463, 251

\bibitem[{{Uttenthaler} {et~al.}(2012){Uttenthaler}, {Schultheis}, {Nataf},
  {Robin}, {Lebzelter}, \& {Chen}}]{2012A&A...546A..57U}
{Uttenthaler}, S., {Schultheis}, M., {Nataf}, D.~M., {et~al.} 2012, \aap, 546,
  A57, {(Paper II)}

\end{thebibliography}

\end{document}